# Role of Symmetry in Irrational Choice

Ivan Kozic

ivan.kozic@tutanota.de

**Abstract**

Symmetry is a fundamental concept in modern physics and other related sciences. Being such a powerful tool, almost all physical theories can be derived from symmetry, and the effectiveness of such an approach is astonishing. Since many physicists do not actually believe that symmetry is a fundamental feature of nature, it seems more likely it is a fundamental feature of human cognition. According to evolutionary psychologists, humans have a sensory bias for symmetry. The unconscious quest for symmetrical patterns has developed as a solution to specific adaptive problems related to survival and reproduction. Therefore, it comes as no surprise that some fundamental concepts in psychology and behavioral economics necessarily involve symmetry. The purpose of this paper is to draw attention to the role of symmetry in decision-making and to illustrate how it can be algebraically operationalized through the use of mathematical group theory.

## I. Introduction

It would not be entirely incorrect to claim that behavioral economics is all about irrational choice. It has been developed to explain the anomalies in rational choice theory, the pillar of modern mainstream economics. By preferring psychological concepts over the outcomes of mathematical models, behavioral economists manage to successfully explain real-world microeconomic phenomena. Although most fundamental notions in behavioral economics have been well explained and thoroughly tested by psychologists, thinking about them on an abstract level necessarily involves the emergence of a distinctive and specific concept that has not been hitherto extensively discussed, either by psychologists or economists. This distinctive and specific concept is symmetry – a powerful deductive tool in modern physics and all related sciences, thereby hinting its usefulness in social sciences as well.

It would also not be entirely incorrect to claim that contemporary physics is all about symmetries. Starting from Einstein's (1905) seminal work, symmetry has become the ultimate tool for developing physical theories. One of the earliest comprehensive books on symmetry is Weyl's (1952). The German mathematician recognized the power of symmetry in geometry, art, and nature. Then, for the first time, some philosophical aspects of symmetry were systematically elaborated and discussed. The idea of the unmistakable approach of "putting symmetry first" has unstoppably evolved to the present day, when, for example, Schwichtenberg (2015) comprehensively demonstrated how all the pillars of modern physics can be derived from symmetry. For instance, the so-called spatiotemporal symmetry implies conservation of the laws of nature under the spatial or temporal displacement of an object under consideration. It is the symmetry required for the derivation of Einstein's special relativity, and forms the basis of physics as related to the macroscopic world. On the other hand, gauge symmetry is required for understanding the microscopic world. Specifically, it forms the basis of the standard model of particle physics, and represents the most important cog in the discovery of the Higgs boson.

Although there are various symmetries in physics, all involve conservation of one or more essential features of the physical state or law under change. Therefore, symmetry essentially means immunity to change. If we prefer to sound more scientific, we would say that symmetry is invariance under transformation.

Many of today's leading physicists do not actually believe symmetry is a fundamental property of nature (Livio 2006), as nature does not seem to be symmetrical at all. Most likely, it is only nearly symmetrical. However, its laws (mostly) exhibit exact symmetry. It should be kept in mind that the laws of nature represent, in fact, our idea of what is happening around us, and their deduction is apparently allowed by the symmetry deeply embedded in our cognition. As such, we can consider symmetry as a rather subjective and intuitive underlying principle of nature. It is "subjective" because it is unlikely that symmetry is a feature of the material world, but more likely embedded in our way of reasoning.

Beyond this general overview, we will now attempt to identify the correct place of symmetry in the world of behavioral economics. The rest of paper is organized as follows. Section II discusses the importance of symmetry for human perception and cognition, and then briefly discusses the possible origin of cognitive symmetry bias. Section III describes the role of symmetry in some fundamental concepts of behavioral economics and particularly analyzes the concept of reference-dependence, widely regarded as the central idea of behavioral economics. Section IV presents group theory as the mathematical language of symmetry, and therefore the natural language of behavioral economics. Section V provides the concluding remarks.

## II. Importance of Symmetry for Human Perception and Cognition

Did you ever consider the repeating of tones on a musical scale? If we were pressing the keys on a piano keyboard one after another, we would hear exactly the same sound at every eighth step. In fact, what we would hear is pitch, an auditory sensation, which is a subjective outcome of our perception. On the other hand, pitch is also the outcome of frequency, produced by the vibration of strings inside the piano, where frequency is an objective part of the physical world. It is measurable and does not repeat on the piano keyboard. The question is, why do we hear exactly the same sound related to completely different frequencies. This is because every repeating tone is the harmonic of a preceding one. The auditory similarities between fundamental tones and their harmonics are determined by the specific anatomies of the associated acoustic waves. A harmonic is, in fact, represented by the same acoustic wave as the fundamental tone, the only difference being the scalar that serves as the multiplication factor appearing in the analytical expression of the harmonic's acoustic wave. Therefore, the auditory similarity is the consequence of an objective feature of the physical world. On the other hand, if we kept running the piano keyboard, we would also notice specific auditory similarities between tones, represented by frequencies that have nothing in common. For instance, we would notice the auditory similarity between each two pairs of tones that form the basic building block of a chord. It is the consequence of the so-called "circle of fifths" phenomena that form the core of western music. The term "fifth" represents the closest interval between two tones showing auditory similarity, which is not a consequence of shared physical attributes. This auditory similarity is the apparent consequence of the symmetry embedded in our minds. Specifically, it is the consequence of the so-called rotational symmetry, which enables the arrangement of tones in scales and chords, thereby enabling the production of gorgeous sound, such as Ravel's Bolero or Nirvana's Smells like Teen Spirit.

Symmetry is a kind of perceptual-cognitive tool for organizing and systematizing large amounts of information from the environment into coherent knowledge. If we would not have symmetry embedded in our perceptual-cognitive apparatus, we could not be able to organize all those different frequencies in a beautifully sounding melody, but would only hear noise. Moreover, with no symmetry, we could not be able to systematize all elementary particles into the standard model of particle physics. Further, without symmetry, we would barely be able to comprehend gravitation as the consequence of the curvature of space-time. There are numerous other examples, but the most intriguing question is where does it actually come from?

Enquist and Arak (1994, p. 169) argue that "humans and some other species have sensory bias for symmetry. Symmetrical patterns are just more attractive to them than asymmetrical ones." These patterns are easily noticeable and quickly comprehensible, which is an important consequence of adaptive behavior in the evolution of these species. According to evolutionary psychologists, symmetry plays an extremely important role in adaptation to environment, thereby enabling survival and reproduction. The bias towards symmetry evolved as a consequence of the need to recognize a predator, identify potential food, or choose a healthy mating partner (Livio 2006). In addition, symmetry detection plays a crucial role in feelings of safety. As such, it comes as no surprise that anxiety disorders very often involve perfectionism and obsessive-compulsive behaviors, and the quest for symmetry links them together. An anxious person keeps performing obsessive-compulsive actions, as long as his/her desire for symmetry is not appropriately satisfied. For instance, an anxious person often keeps arranging things on a bookshelf until they are perfectly aligned. Here, symmetry plays the role of a "norm," as the action of arranging things is repeated until the "norm" is finally met. This makes the anxious person more comfortable and anxiety decreases. Not only are mental disorders accompanied by strong preferences for symmetry. "Nearly 90% of the general population report some obsessions, preoccupations or compulsive urges involving symmetry" (Evans et al. 2012, p. 2).

As a result, it could be concluded that symmetry is naturally inherent in human perception, cognition, and, consequently, decision-making. The symmetry bias was useful in the past when humans needed to avoid predators and recognize food. On the other hand, the preferences for symmetry are presently useful when symmetry serves as a cognitive tool for the organization and systematization of information. However, problems occur when the human mind tries to detect symmetry even where there is none. Symmetry stops then to be useful and, instead, leads to severe and systematic errors in decision-making. To clarify this, we need to introduce a more familiar psychological concept – analogy.

According to Hofstadter and Sander (2013), analogy is the core of cognition. It is, in fact, the fuel and fire of thinking. They argue that analogy plays a central role in concept production and evocation and depict it as a mental phenomenon of selective exploitation of past experiences to clarify new and unfamiliar concepts or ideas. Encyclopedia Britannica describes analogy as a type of reasoning that presents similarities, while Wikipedia defines analogy as a cognitive process of transferring information from a particular object to another one. From the abstract level, an analogy is in fact symmetry (Rosen 2008). This logically leads to the conclusion that analogy/symmetry is responsible for the reference-dependence in human thinking and decision-making. Therefore, as the rest of this paper shows, analogy/symmetry forms the kernel of some important concepts in behavioral economics.

## III. Fundamentals of Behavioral Economics: Revelation of Symmetry

To the best of my knowledge, one of the most cited papers in behavioral economics is Kahneman and Tversky (1979), which proposed prospect theory, the pillar of contemporary behavioral economics, thereby paving its entrance into the mainstream of modern economic thought. The paper was preceded by another important study, Tversky and Kahneman (1974). Although this earlier piece is not as well known, it has introduced the critical background subsequently used in the development of prospect theory. Tversky and Kahneman (1974) described three essential heuristics that lead to severe biases in human thinking as follows.

First, according to Tversky and Kahneman (1974, p. 1124), representativeness is a simple heuristic "in which probabilities that process B will generate event A are evaluated by the degree to which A is representative of B, that is, by the degree to which A resembles B." Since "resemblance" specifically means "similarity," it is not difficult to conclude that representativeness, in fact, implies analogy in its purest sense. The mental application of the representativeness heuristic between events A and B represents the process of drawing an analogy between A and B, thereby introducing symmetry into this action.

Second, availability is a heuristic, whereby "people assess the frequency of a class or the probability of an event by the ease with which instances or occurrences can be brought to mind" (Tversky and Kahneman 1974, p. 1127). The heuristic of availability implies a process of drawing an analogy between the event and something that can be easily retrieved from memory. The effortlessness of symmetry detection plays a crucial role in the utilization of this heuristic.

Finally, anchoring is a heuristic, whereby "people make estimates by starting from an initial value that is adjusted to yield the final answer. The initial value, or starting point, may be suggested by the formulation of the problem, or it may be the result of a partial computation" (Tversky and Kahneman 1974, p. 1128). Anchoring represents the process of drawing an analogy between a starting point and the final result.

Although it has not been explicitly noted by Tversky and Kahneman (1974), the three essential heuristics are, in fact, closely related to the concept of reference-dependence, which is a type of cognitive processing whereby individuals tend to assess the consequences of their decisions relative to a reference point. A careful analysis can reveal that the concept of reference-dependence involves two hidden cognitive levels: the unconscious level implies drawing an analogy between the potential outcome of a decision and the reference point, while the abstract level is the attempt towards symmetry detection.

In her doctoral dissertation, German economist Evelyn Stommel (2013) conducted a theoretical and experimental investigation of what she considers to be an essential topic in behavioral economics – formation of reference-dependent preferences. As Stommel (2013) suggests, Kahneman and Tversky (1979), Thaler (1980), and Tversky and Kahneman (1991) have, in fact, shown how changes from reference points form the foundation of human behavior. According to Kahneman and Tversky (1979), the essential feature of their theory is indeed the change from a reference point. To this end, they have proposed the value function, defined by deviations from the reference point. Further, Thaler (1980) tried to connect it with the theory of consumer choice, while Tversky and Kahneman (1991) finally proposed a reference-dependent theory of consumer choice.

Moreover, Ariely, Loewenstein, and Prelec (2003) have shown that reference-dependence does not need to have a rational basis. As confirmed by their experiments, anchors (i.e., reference points) could be chosen totally arbitrarily. For example, a decision-maker's social security number could serve as his/her reference point in decisions on the price he/she is willing to pay for a consumer good. This appears nonsensical, unless analogy/symmetry is considered. Since the decision-maker tries to draw an analogy between an unknown price and something familiar, his/her irrational mistake seems to be an inevitable consequence of the unavailability of suitable reference points. He/she then tends to establish symmetry between something to decide about and something at hand. His/her desire for symmetry is so strong that he/she is not aware of doing something completely irrational – applying the transformation of the social security number into the price. The series of behavioral experiments presented by Ariely (2008) show that such mistakes are rather frequent, and represent systematic features of human decision-making.

Another typical example is the phenomenon of loss aversion, accompanied by the endowment effect stating that decision-makers tend to value owned goods more than prior to owning them. In terms of symmetry, a decision-maker, facing the decision moment, is actually projecting his/her own status on the future by drawing an analogy, that is, seeking symmetry between current and future statuses. Since he/she would prefer keeping future status symmetrical to the reference point (current status), he/she is ready to pay for that. However, overvaluation implies the loss of opportunity to obtain money and more choices (to buy similar or other goods). The decision-maker unconsciously accepts this fact due to his/her desire for symmetry.

It comes as no surprise that symmetry seeking reminds us of status quo bias. In fact, status quo bias represents the most obvious example of symmetry at work: decision-makers are rather reluctant to change their current status despite all opportunities they miss. This phenomenon is identified in many different decision-making situations, and all of them involve the action of drawing analogies between the current and possible future statuses. In fact, this is the attempt to detect symmetry between the reference point and the decision outcome. Since humans feel comfortable and safe upon detecting symmetry, it comes as no surprise they tend to maintain the current status. They actually want to make their future status symmetrical to the current one.

This context reminds me of the situation of one of my female friends. She always tends to fall in love with the "wrong guy" (whom she calls "the same kind of bastard"). The origin of her problem is, in fact, the symmetry deeply embedded in her mind: her first boyfriend serves as the reference point. It seems as if she always tries to detect symmetry between her first boyfriend and all others. She unconsciously picks the ones that have most in common with her first boyfriend. In other words, she draws an analogy, that is, seeks symmetry and thus unmistakably commits the same error she always tries to avoid. She finally ends up with the "same guy" repeatedly.

As shown by these examples, the quest for symmetry represents the underlying mechanism that sometimes leads to severe and systematic errors in decision-making, one of the main reasons we are irrational. To express this in economics terms, symmetry plays a crucial role in the formation of reference-dependent preferences and the related economic phenomena. To clarify that, we need to introduce the symmetry principle (sometimes called Curie's principle after the French physicist Pierre Curie), which states that any symmetry of a cause must appear in its effect. Therefore, symmetry being responsible for the formation of reference-dependent preferences necessarily appears in all economic phenomena, arising out of reference-dependent preference formation. Consequently, utility, demand, and price necessarily contain the same symmetry as the preferences they are derived from. It is a very important fact which simplifies economic analysis. We can now be sure that everything in an economy could be driven by the elementary symmetry responsible for reference-dependent preference formation. Unfortunately, everything could also be driven by error accidently committed at the moment of preference formation. The rest of the paper employs mathematical group theory to illustrate the permeation of symmetry into the entire economy.

## IV. Group Theory as the Language of Symmetry

As already noted in the introduction, symmetry represents the immunity to change (i.e., invariance in transformation). To make this statement mathematically operational, we need to introduce mathematical group theory. Although group theory has been independently developed, it is presented here in close relation to symmetry analysis. Therefore, the essential definitions of group theory are adjusted to fit symmetry analysis without the need for further explanations.

Let us start with the basic definitions. A transformation is an action of mapping from an initial state to a final one, and can be active or passive. Active transformation is the transformation of an object under consideration, holding the reference frame unchanged, while passive transformation is the transformation of the reference frame.

A transformation group is a set of transformations ($G$), including the binary operation that combines any two transformations to form a composite transformation (the binary operation represents, in fact, the consecutive application of transformations). To be a group, such a set must satisfy the following requirements:

1) <u>Existence of identity element</u>

It is a neutral transformation ($i \in G$), that is, a transformation without any impact:

$$\text{If } a \in G, \text{ then } ia = ai = a$$

2) <u>Existence of inverse element</u>

It is a specific transformation ($a^{-1} \in G$) that cancels the action made by its counterpart:

$$\text{If } a \in G, \text{ then there is } a^{-1} \text{ such that } aa^{-1} = a^{-1}a = i$$

3) <u>Associativity</u>

The arrangement of application pairs of transformations is irrelevant. The order could not impact the final result:

$$\text{If } a, b, c \in G, \text{ then } a(bc) = (ab)c$$

4) Closure

The mutual application of transformations yields another (composite) transformation belonging to the same group:

$$\text{If } a, b \in G, \text{ then } ab \in G$$

A symmetry group is a subgroup of the transformation group. It contains only those transformations that leave the object under consideration (its physical or monetary attributes) unchanged.

A group representation is a description of the elements of a transformation/symmetry group in terms of more familiar and intuitive mathematical objects. For instance, a group of abstract transformations can be represented by a group of linear algebra objects, whereby every element of the transformation group gets a linear algebra counterpart in such a way that all properties of the transformation group remain preserved. Therefore, representation theory is crucial for making the abstract transformations mathematically operational. It also enables the description of abstract transformations by the use of linear algebra artifacts. As such, any abstract problem of group theory can be reduced to a well-known linear algebra problem. In fact, representation theory is making symmetry analysis mathematically feasible.

Let us demonstrate the possible application of symmetry analysis in the context of behavioral economics. The following simple example shows how symmetry analysis can be used to connect the concept of reference-dependence with standard economic concepts such as utility, demand, and price.

Let us assume that $x$, $y$, and $z$ represent one monetary and two physical attributes of an object $R$, which serves as the reference point in the process of reference-dependent preferences formation. $\alpha$, $\beta$, and $\gamma$ represent transformation parameters. The result of the transformation application is as follows:

$$\alpha x, \beta y, \gamma z$$

For simplicity, the parameters ($\alpha$, $\beta$, $\gamma$) of three active transformation are replaced by a single parameter ($\varepsilon$) of one passive transformation:

$$\varepsilon R$$

For the transformation to be symmetry transformation, the physical and monetary attributes of the object $R$ have to remain unchanged:

$$\varepsilon R = \mathrm{R} \Leftrightarrow \alpha x = x, \beta y = y, \gamma z = z$$

Reference-dependence involves the inclusion of the symmetry transformation in the preference structure:

$$A \succ B \succ C \Rightarrow A \equiv \varepsilon R$$

Consequently, the applied transformation results in the inclusion of symmetry ($A \equiv \varepsilon R = R$) in all economic concepts that are consecutively derived from the newly formed preferences.

Therefore, the representing utility function takes the following reference-dependent form:

$$u(R) > u(B) > u(C)$$

The symmetry impacts the derived demand function:

$$x_A = f(p_R, m)$$

Finally, it also impacts the price willing to be paid on the market:

$$p_A = f(x_R)$$

The symmetry is apparently responsible for the price being finally influenced by object *R*, which served as the reference point at the beginning of the process. Although the resulting analytic formulation is identical to the standard expression used for the description of reference-dependence in modern economic literature, the benefits of this approach could be derived from the potential of group theory.

The genuine beauty of group theory is its potential to reveal the properties of an object by analyzing symmetry transformations without observing the object itself. Moreover, if different objects permit the same symmetry transformations, group theory allows us to analyze them simultaneously. Finally, it allows us to recognize hidden similarities between seemingly different objects (Mirman 1995). Group theory can be thus employed to observe known economic problems from a completely new viewpoint. It could potentially lead to the discovery of some relations that remain hidden under the application of traditional analytical tools. There could be many different systems having the same symmetries, thereby being described by the same group. Accordingly, group theory could provide a narrow list of possible behaviors for those systems (Mirman 1995).

Some practical examples of group theory application in the context of mainstream economics are presented by Sato and Ramachandran (2014). Although they are rather instructive, these examples do not consider the fundamentals of behavioral economics, which imply that symmetry is an important driver of reference-dependent preferences formation, thereby logically promoting group theory as the natural language of behavioral economics. An illustration of proper group theory application with reference to the fundamentals of behavioral economics is presented here. However, the technical details and real case examples are far beyond the scope of this paper. They are topics that need to be discussed and explored in future studies.

## V. Concluding Remarks

Humans' ability to understand natural and social phenomena is determined by their perceptual-cognitive apparatus, in which symmetry plays an important role. As such, it is reasonable to question whether our comprehension of complex phenomena, like quantum mechanics or social/economic interactions, is correct. Although we can effectively manage the means and outputs of those processes, we still have doubts on what is really happening at a more profound level. Our comprehension capacity is simply reduced and constrained by symmetry, deeply embedded in our cognition. Although this trait was an advantage when we had to survive nature, it could be considered a critical disadvantage at present, when nature has to survive us. Therefore, it is important to be aware of situations where the desire for symmetry could lead to irrational choices.

The primary aim of this paper is to draw attention to the role of symmetry in human decision-making, thereby revealing the huge potential of symmetry analysis in the field of behavioral economics. It is very likely a hidden underlying principle – the quest for symmetry – found in the background of human cognition. It determines our ability to understand what is happening around us and sometimes leads us to serious biases that could turn into systematic errors in our comprehension and decision-making. This underlying principle is waiting to be explored in the context of behavioral economics, thereby paving the way for the application of symmetry analysis as an effective tool for analyzing human behavior in social and economic interactions.

Following the Great Recession, orthodox economics is still under attack. Meanwhile, behavioral economics discloses the fundamental weaknesses of rational choice theory. At the same time, neuroeconomics remains on its own path of setting the foundations of behavioral economics. These foundations might easily turn behavioral economics into a more exact discipline than orthodox economics could ever be. Since the quest for symmetry seems to be an integral part of human cognition, symmetry analysis and, therefore, group theory should find their places in the future development of behavioral economics. Group theory could thus become the mathematical language of behavioral economics. As calculus seems to be somewhat deficient in attempting to employ the full potential of psychological concepts that constitute behavioral economics, it is very likely that group theory could emerge as an elegant and convenient solution. It could finally appear that symmetry analysis (group theory) could be the missing link in merging psychology and economics into an integral and comprehensive social science.

## REFERENCES

Ariely, D. (2008) Predictably Irrational: The Hidden Forces That Shape Our Decisions. New York: HarperCollins.

Ariely, D., Loewenstein, G., Prelec, D. (2003) Coherent Arbitrariness: Stable Demand Curve Without Stable Preferences, The Quarterly Journal of Economics 118 (1), 73-106.

Einstein, A. (1905) Zur Elektrodynamik bewegter Körper, Annalen der Physik 322 (10), 891-921.

Enquist, M., Arak, A. (1994) Symmetry, Beauty and Evolution, Nature 372, 169-172.

Evans, D. W., Orr, P. T., Lazar, S. M., Breton, D., Gerard, J. et al. (2012) Human Preferences for Symmetry: Subjective Experience, Cognitive Conflict and Cortical Brain Activity, PLoS ONE 7 (6), doi: 10.1371/journal.pone.0038966.

Hofstadter, D., Sander, E. (2013) Surfaces and Essences: Analogy as the Fuel and Fire of Thinking. New York: Basic Books.

Kahneman, D., Tversky, A. (1979) Prospect Theory: An Analysis of Decision under Risk, Econometrica 47 (2), 263-292.

Livio, M. (2006) The Equation That Couldn't Be Solved: How Mathematical Genius Discovered the Language of Symmetry. New York: Simon & Schuster.

Mirman, R. (1995) Group Theory: An Intuitive Approach. Singapore: World Scientific.

Rosen, J. (2008) Symmetry Rules: How Science and Nature Are Founded on Symmetry. Berlin: Springer-Verlag.

Sato, R., Ramachandran, R. V. (2014) Symmetry and Economic Invariance. New York: Springer.

Schwichtenberg, J. (2015) Physics from Symmetry. New York: Springer.

Stommel, E. (2013) Reference-Dependent Preferences: A Theoretical and Experimental Investigation of Individual Reference-Point Formation. Wiesbaden: Springer Gabler.

Thaler, R. (1980) Toward a Positive Theory of Consumer Choice, Journal of Economic Behavior and Organization 1 (1), 39-60.

Tversky, A., Kahneman, D. (1974) Judgement under Uncertainty: Heuristics and Biases, Science 185 (4157), 1124-1131.

Tversky, A., Kahneman, D. (1991) Loss Aversion in Riskless Choice: A Reference-Dependent Model, The Quarterly Journal of Economics 106 (4), 1039-1061.

Weyl, H. (1952) Symmetry. Princeton: Princeton University Press.